\documentclass[pra, amsfonts, amssymb, amsmath, reprint, showkeys, twoside, nofootinbib, aps, superscriptaddress, print, longbibliography]{revtex4-2}
\usepackage{pgfplots,graphicx,caption,float,mathrsfs,placeins,balance,dsfont,braket,balance, amsthm}
\newtheorem{theorem}{Theorem}[section]
\newtheorem{lemma}[theorem]{Lemma}

\definecolor{steelblue}{RGB}{0, 51, 97}

\newcommand{\ee}{\mathrm{e}}
\newcommand{\ii}{\mathrm{i}}
\newcommand{\mcalI}{\mathcal I_{\mathrm s}}

\usepackage[unicode=true,
 bookmarks=false,
 breaklinks=false, colorlinks=true]
 {hyperref}
\hypersetup{
  linkcolor = steelblue,
  citecolor = steelblue,
  urlcolor = steelblue,
}

\begin{document}
\title{Hierarchical Fusion Method for Scalable Quantum Eigenstate Preparation}

\author{Matthew Patkowski}
\affiliation{JILA and Department of Physics, University of Colorado Boulder, Boulder CO 80309}
\author{Onat Ayyildiz}
\affiliation{Facility for Rare Isotope Beams, Michigan State University, East Lansing MI 48824}
\affiliation{Department of Physics \& Astronomy, Michigan State University, East Lansing MI 48824}
\author{Matja\v{z} Kebri\v{c}}
\affiliation{JILA and Department of Physics, University of Colorado Boulder, Boulder CO 80309}
\author{Katharine L. C. Hunt}
\affiliation{Department of Chemistry, Michigan State University, East Lansing MI 48824}
\author{Dean Lee}
\affiliation{Facility for Rare Isotope Beams, Michigan State University, East Lansing MI 48824}
\affiliation{Department of Physics \& Astronomy, Michigan State University, East Lansing MI 48824}

\begin{abstract}
Robust and efficient eigenstate preparation is a central challenge in quantum simulation in the noisy intermediate-scale quantum (NISQ) era and beyond. The Rodeo Algorithm (RA) \cite{choi2021rodeo} offers exponential convergence to a target eigenstate but suffers from poor performance when the initial state has low fidelity with the desired eigenstate, limiting its scalability.
In this work, we introduce a \textit{fusion method} that preconditions the RA by an adiabatic ramp and modular subsystem fusion, enabling high-fidelity state preparation across large many-body systems.
Numerical simulations of the spin-1/2 XX chain demonstrate that this hybrid preconditioning restores the RA's exponential convergence and dramatically reduces computational cost. 
The method's efficiency is governed by boundary size, making it ideally suited for 1D and quasi-1D architectures such as trapped-ion chains and neutral-atom arrays.
The fusion method thus defines a scalable framework for high-fidelity quantum state preparation across both present and future generations of quantum hardware.
\end{abstract}

\maketitle
\balance

\section{INTRODUCTION}
Preparing specific eigenstates of quantum many-body systems is a foundational task across quantum computing, with applications to quantum chemistry, condensed matter physics, and numerous other fields that require simulations of quantum matter. 
Classical simulations of such quantum systems---ranging from exact diagonalization techniques to more advanced approximation methods---are constrained to handling systems of only a few tens of qubits in the former case \cite{dagotto1994correlated, sandvik2010computational}, and at most a few hundred qubits for the latter \cite{georges1996dynamical, troyer2005computational}.
Quantum computers overcome this challenge but have been limited for past decades by noise and small system size. However, modern quantum platforms have demonstrated rapid advances in coherence times, gate fidelities, and qubit scalability \cite{siddiqi2021engineering, moses2023race, bruzewicz2019trapped, omran2019generation, browaeys2020many}. Moreover, with recent developments in robust quantum error correction techniques \cite{campbell2024series, ai2024quantum} and progress toward error-robust architectures \cite{microsoft2025interferometric}, quantum computers capable of outperforming classical methods are becoming increasingly plausible.
Thus, it is of particular interest to develop scalable quantum computing algorithms. 

In this work, we improve upon the recently introduced Rodeo Algorithm \cite{choi2021rodeo}, which has been shown to outperform traditional eigenstate preparation methods such as quantum phase estimation \cite{kitaev1995quantum} and adiabatic quantum computation \cite{rezakhani2010accuracy, albash2018adiabatic}. Its practical robustness has been demonstrated on contemporary quantum hardware \cite{qian2024demonstration}, and several enhancements have been proposed to further improve its performance, including a reduction in entangling gate complexity through the use of controlled reversal operations \cite{bee2024controlled}, and a constant-factor speedup enabled by incorporating spectral information of the underlying Hamiltonian \cite{cohen2023optimizing}. However, the primary limitation of the Rodeo Algorithm remains its inefficiency when the initial state's overlap with the target eigenstate is low---a challenge that becomes increasingly significant with system size. To address this, we propose a hybrid ``fusion" approach that first employs an adiabatic ramp to precondition the input state of the Rodeo Algorithm. Additionally, we demonstrate a modular construction scheme wherein the full system is assembled from smaller subsystems, each prepared with high fidelity, enabling scalable eigenstate preparation in large quantum systems.

Moreover, the fusion method constitutes a concrete example in a broader design philosophy of quantum algorithms, of hybrid methods leveraging preconditioned approximate solutions as inputs to a fast refinement stage.
In the particular construction we present, the fusion method is highly efficient for one-dimensional and quasi-one-dimensional systems where the added boundary remains small. Similarly inspired algorithms may use alternate lattice construction methods to extend to higher-dimensional systems. Nevertheless, many experimental simulators, Rydberg atom arrays in particular, with linear or ladder geometries can prepare high-quality states at tens of qubits \cite{chen2024benchmarking, zhang2025probing, endres2016atom}. Recent work has demonstrated fast and high-fidelity entanglement generation capabilities between remote subsystems in $^{171}\mathrm{Yb}$ neutral-atom processors via zone-based shuttling and modular interconnects, providing a promising platform for scalable quantum algorithms \cite{li2024high}. These platforms are also demonstrated to support fault-tolerant computation with high-fidelity gates \cite{wu2022erasure, ma2023high}. In this regard, the fusion method is of immediate and practical applicability to such geometries in NISQ devices, where minimizing circuit complexity and runtime is essential, as well as to post-NISQ, fault-tolerant devices that require scalable algorithms.

\section{METHODS}
\begin{figure*}[t]
    \centering
    \includegraphics[width=\linewidth]{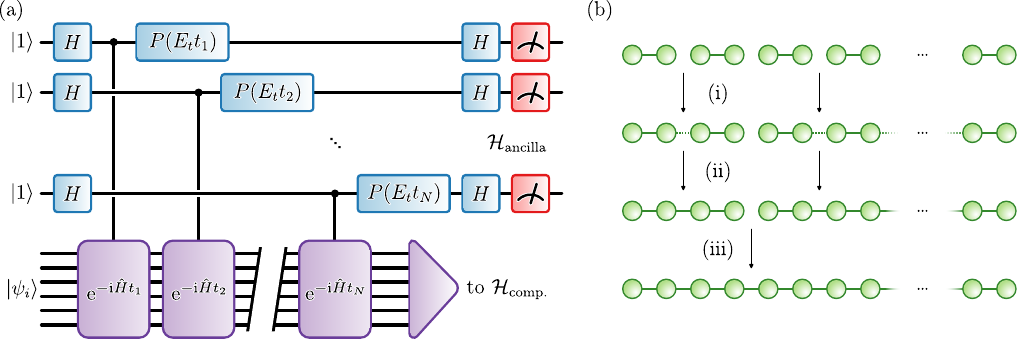}
    \caption{
    (a) Rodeo Algorithm circuit. Ancilla qubits are initialized in $|1\rangle$ and are used as control qubits for controlled time evolution. Each ancilla corresponds to a time sample $t_j$. A phase kickback from the controlled unitaries is propagated through the ancillas and a successful Rodeo Algorithm application is categorized by measurement of all $|1\rangle$ states on the ancillas after the algorithm \cite{choi2021rodeo}. Specifically, if $|\psi _i\rangle$ is an eigenstate of $\hat H$ with energy $E_t,$ the phase kickback from the unitary will be exactly canceled by the phase gate, such that the ancilla always measures one. Contrastingly, when $|\psi _i\rangle$ is not the exact targeted eigenstate, the orthogonal components of the input state will cause phase kickbacks that do not get canceled by the phase gate, and lead to a decrease in the success probability. (b) Illustration of the fusion approach. Instead of directly preparing a many-body quantum state, we subdivide the system, in this case, in a binary fashion. Each ``fusion step" consists of an (i) adiabatic preconditioning and the (ii) RA purification. (iii) Fusion steps are repeated until the full system is reconstructed.
    }
    \label{fig:combinedrodeoandfusion}
\end{figure*}
The Rodeo Algorithm (RA) \cite{choi2021rodeo} (see Fig.~\ref{fig:combinedrodeoandfusion}) is a circuit-based stochastic quantum computing algorithm that prepares a ``target" eigenstate $|E_t\rangle$ of given Hamiltonian $\hat H$ through controlled time evolution. Given this Hamiltonian exists on system of computational qubits $\mathcal H_{\text{comp.}}$, one requires an ancillary system of qubits $\mathcal H_{\text{ancilla}}$ which undergo phase kickback and encode a successful run of the cycle. One specifies the energy of the target eigenstate $E_t$ and encodes an initial guess $|\psi _i\rangle$ on the ensemble of computational qubits. One then samples times $t_1, t_2, \ldots, t_N$ for the controlled time evolution and phase gates as in Fig.~\ref{fig:combinedrodeoandfusion}.
 Measuring all ancilla qubits in the $|1\rangle$ state indicates a successful RA application.
The post-rodeo state, conditioned on all ancillas measured as $|1\rangle$, may be analytically derived (see Appendix \ref{appen:analyticalra}): \begin{equation}|\psi \rangle = \frac{1}{2^N}\prod _{j=1}^N \left(\hat I + \ee^{-\ii(\hat H -E_t)t_j}\right) |\psi _i\rangle.\end{equation} Note the non-unitarity of the above action, as this state is conditioned on measuring $|1\rangle ^{\otimes N}$ on the ancilla subsystem, the success probability of the RA is $\langle \psi |\psi \rangle.$ 

Note the action of the RA is constructed such that if $E_t$ is perfectly selected on an eigenstate of $\hat H,$ the spectral weight of that target eigenstate will remain constant in each RA cycle. All other eigenstates will be suppressed by some factor. Thus, in the limit $N\to \infty$ we will have $|\psi \rangle = \alpha |E_t\rangle,$ where $\alpha$ is the initial spectral weight of the target state in $|\psi _i\rangle.$ Evidently, without careful selection of an initial state, the input state will have vanishing overlap with the target state as system size is scaled and the RA fails to scale past a few-qubit system. 

We thus apply a ``binary fusion" approach, building a many-body one-dimensional system from smaller blocks that are easier to prepare, as illustrated in Fig.~\ref{fig:combinedrodeoandfusion}. 
Each ``fusion step" consists of a linear adiabatic ramp from two disconnected systems of size $L /2$ to a single system of size $L$, followed by an application of the RA.
We suggest that a system of arbitrary size can be constructed in this fashion, using a binary-like decomposition of the system size. We assume it is easy to prepare the ground state of the small Hamiltonians. An artifact of this construction is that the long-wavelength (IR) behavior is modified. Hence states with nontrivial IR behavior may require alternate lattice construction methods. 

We demonstrate the fusion method with a numerical simulation of the nearest-neighbor spin-1/2 XX model with Hamiltonian \begin{equation}
    \hat H = 2J \sum _{\langle i,j\rangle} \left(\hat \sigma _i^+ \hat \sigma_j ^- + \hat \sigma _i ^- \hat \sigma _j^+ \right) , \label{eq:1}
\end{equation} where $\hat \sigma_i ^\pm = \frac 12 \sigma _i ^x \pm \frac \ii2 \sigma _i ^y$ and we take units $\hbar = 1$. It follows directly that we have a global U(1) symmetry that restricts dynamics between sectors of given magnetization $\left\langle \sum _i\sigma _i ^ z \right\rangle$. Notably, in this phase, the XX chain is gapless and corresponds to a free fermion model \cite{lieb1961two}. The adiabatic ramp is performed by linearly ramping the interaction strength $J$ of the bond $N\leftrightarrow N+1$ linking the two halves of the system of size $2N$. 
\section{RESULTS}
\begin{figure}[t!]
\centering
    \includegraphics[width=\linewidth]{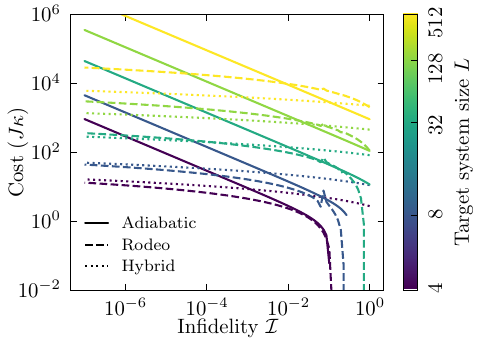}
    \raggedright
    \captionsetup{justification=raggedright, singlelinecheck=false}
    \caption{Method costs for preparing the ground state of an XX model of the indicated size and infidelity.
    Line color represents the merged system size $L \in \{4,8,32,128,512\}$.
    Line style indicates the method: solid lines for pure adiabatic evolution (method 1), dashed lines for the unmodified rodeo algorithm (method 2), and dotted lines for the hybrid method (method 3). 
    }
    \label{figure:xx_results}
\end{figure}
We compare three fusion methods for the problem of preparing the ground state of the complete system of size $L$ beginning with ground states of the 2-qubit XX chain and iteratively applying fusion steps as depicted in Fig.~\ref{fig:combinedrodeoandfusion}b. The methods are: (1) pure adiabatic evolution, (2) pure RA evolution, and (3) a hybrid approach that uses RA evolution starting from a state preconditioned via adiabatic evolution. To evaluate the performance of each method, we define a computational cost $\kappa$ representing the expected total unitary evolution time, penalizing by the expected number of repetitions due to RA failures. To ensure dimensional consistency in calculating expected failure cascades, this continuous time is strictly modeled in multiples of $\tau$, the duration of a single underlying preparation attempt. For a detailed definition and discussion of the cost metric, see Appendix~\ref{appen:costmetric}. 
We compare the costs necessary to reach a given infidelity, defined as $\mathcal I = 1-|\langle \psi |E_t\rangle |^2,$ where $|\psi\rangle$ is the state we obtain from our algorithm.

We present our results in Fig.~\ref{figure:xx_results}, showing the results in the zero magnetization $\langle \sum _i \hat \sigma _i ^z \rangle= 0$ sector (half-filling in the boson description). We assume perfect knowledge of the ground and first excited state energies in the given sector, such that we may input $E_t = E_0$. Time samples are selected in a geometric series of common ratio $\alpha^{-1}$ where $\alpha = 1+\alpha _0 T_{\mathrm r}^{-1}L^{-1}$, where $\alpha_0$ is a fitting parameter that is optimized for small systems of size $L=4,8$. These time samples have been demonstrated to perform near-optimally as the relation between subsequent time samples ensures that individual poor time samples (those that fail to suppress a large part of the undesired spectrum) are compensated by efficient ones  \cite{cohen2023optimizing, patkowski2026improved}. The results for the hybrid method are obtained by preconditioning the input state adiabatically up to an infidelity of approximately $10^{-2}$, as this provides an input guess to the RA that is unlikely to fail. All results are obtained with a free fermion mapping of the system, with scaling laws fitted to the raw data. The adiabatic data is fit with a modified power-law slightly overestimates the performance, while the RA and hybrid methods are fit with an exponential that slightly underestimates the performance, but is sufficient to demonstrate the superiority of the hybrid method. See Appendix~\ref{appen:freefermionsim} for more details.

As demonstrated in Fig.~\ref{figure:xx_results}, continuing to use adiabatic evolution to prepare small infidelities $ \mathcal I\lesssim 10^{-3}$ results in a fast spike in computational cost, reflecting the inherent slowness of the power-law suppression offered by an adiabatic ramp. Contrastingly, the hybrid method results exhibit a rapid convergence across system sizes and infidelities. Furthermore, the figure demonstrates that while using the non-preconditioned RA is relatively effective for smaller system sizes, the decreasing overlap of the non-preconditioned guess with the ground state as the system scales starts to increase the cost through RA repetition penalty, which is particularly evident in the results for $L=512$, where the hybrid method saves an order of magnitude in cost.

In this specific system, the initial ground state of the unfused chain and the target ground state already has modest fidelity ($\gtrsim  0.4$ for the system sizes we simulated). Thus, the payoff in preconditioning adiabatically is not immense. This is, in part, due to the fusion approach, which makes the input to the next fusion step the ground state of a very similar system, making it unlikely that we need to repeat much. However, a small perturbation to the system---in this system, the removal of the middle bond---will, in general, have a greater effect the greater the system size due to the Orthogonality Catastrophe \cite{meden1998orthogonality}. Notwithstanding, the necessity of adiabatic preconditioning is already apparent from the medium-scale results of $L\gtrsim 128$ where the number of consecutive fusion steps that must all succeed incurs a substantial penalty on the non-preconditioned method.

As noted, the results assume that we have perfect knowledge of the ground state energy $E_0$, which may not be readily available; however, as demonstrated in \cite{choi2021rodeo, qian2024demonstration}, one can employ a sweep over $E_t$ to determine the spectrum. Although this adds an overhead to the runtime, we comment that such a sweep need only be done once for a system, and with smartly-chosen energy bounds on the ground state, one can reduce the search area. After determining this energy level, the RA readily and efficiently initializes the eigenstate without need for a sweep over energies. 

\section{DISCUSSION AND OUTLOOK}
We have demonstrated an extension of the RA to large systems where it would normally suffer in performance due to poor initial states. This is achieved by two insights: (1) the binary fusion procedure, involving splitting a large system into more tractable subsystems and (2) the hybrid method, adiabatically preconditioning the input state to the RA. 
For the one-dimensional XX model considered here, and more generally for systems on linear or quasi-linear geometries, this strategy is especially natural as such systems can be recursively partitioned into smaller blocks whose ground states already share moderate fidelity with the fused target state.
As the system size increases, the preconditioned hybrid approach increasingly outperforms the unmodified RA, reducing the repetition cost associated with poor initial overlap. In this work, we have focused on the case where the ground state does not exhibit phase separation.  For systems with phase separation, such as the formation of self-bound droplets, the fusion method can be applied with an external potential in order to localize the different phases to different spatial locations.

The fusion method is efficient insofar as the input state resembles the target state to some degree. In the particular construction we present, partitioning the lattice into smaller sections and ``fusing" it into larger sections modifies the IR behavior. Other hybrid methods may be designed to instead preserve the IR behavior at the expense of accurately reproducing the UV behavior. Additionally, the performance rests on the fusion boundary and entanglement across it remaining small, hence being most relevant for 1-dimensional and quasi-one-dimensional systems. In this regard, the applicability of the fusion method aligns naturally with that of tensor networks which perform optimally in systems with limited entanglement growth.
\begin{figure}[t]
    \centering
    \includegraphics[width=\linewidth]{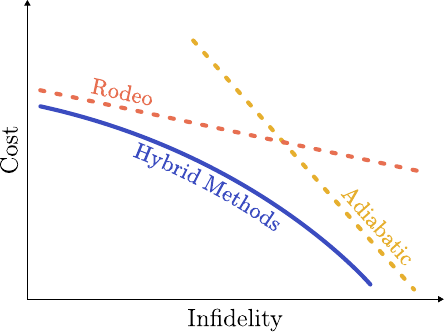}
    \caption{General scheme of the performance of hybrid methods.}
    \label{fig:hybrid_ill}
\end{figure}
The fusion method is one instance of a hybrid quantum algorithm in which a slow preconditioning step prepares a well-conditioned input state for a purification step. From the XX chain results, we posit that such hybrid methods will have a similar cost versus infidelity as seen in Fig.~\ref{fig:hybrid_ill}, where the hybrid method will first have a power-law decrease in infidelity, similar to adiabatic, but once it is well-preconditioned it will speed up to exponentially converging similar to the employed purification algorithm.

Recent developments in hybrid quantum algorithms have focused heavily on subspace methods, such as Quantum Krylov Diagonalization \cite{zhang2024measurement}, which mitigate noise by projecting the problem into a lower-dimensional subspace. While effective, these global approaches often struggle with the orthogonality catastrophe in larger systems without deep circuit precursors. In contrast, our fusion method leverages the emergence of modular quantum architectures, particularly in neutral atom arrays where zone-based shuttling and modular interconnects have been recently demonstrated \cite{li2024high}. By aligning the algorithm's structure with these hardware capabilities---physically constructing the many-body state from smaller, high-fidelity subsystems---the fusion method circumvents the exponential scaling costs associated with preparing large-scale entangled states from scratch. Other hybrid methods including \cite{tabares2025estimating} only require global control and do not need controlled-gate operations, but are limited to computing the ground state energy or simple order parameters rather than computing full spectra or preparing eigenstates such as the Rodeo Algorithm. Nevertheless, such hybrid methods exhibit very similar scaling to the fusion method, with low-cost performance improvement being only marginal to adiabatic methods, followed by a quick improvement in cost after a certain amount of allotted resources.

Together, these results suggest that the fusion method, and more broadly hybrid algorithms, are generally powerful strategies for scalable quantum simulation across both NISQ and post-NISQ platforms.

\section{ACKNOWLEDGMENTS}
We are grateful for discussions with Nathan Jansen.  This research is supported in part by U.S. Department of Energy (DOE) Office of Science grants DE-SC0023658, DE-SC0024586, DE-SC0026198, and DE-SC0013365 and U.S. National Science Foundation (NSF) grant PHY-2310620. 
This work has also been supported in part by NSF grant CHE-2154028 to KLCH.
\balance
\clearpage
\onecolumngrid
\appendix
\section{Post-Rodeo Algorithm State}
\label{appen:analyticalra}
Here we derive the expression for the post-Rodeo Algorithm state analytically. Consider a single RA run with time sample $t_1$ and target energy $E_t$. Then, by the structure of the circuit in Fig.~\ref{fig:combinedrodeoandfusion}, the post-RA state in the basis $\mathcal H_{\text{ancilla}} \otimes \mathcal H_{\text{computation}}$ is given as \begin{align}|\text{ancilla} \rangle \otimes |\psi \rangle &= \frac 12  |0\rangle \otimes (\hat {I} - \exp[-\ii(\hat H - E_t)t_1])|\psi _i\rangle  
        + \underbrace{\frac 12 |1\rangle \otimes (\hat {I} + \exp[-\ii(\hat H - E_t)t_1])|\psi _i \rangle  }_{\text{successful run}}.
\end{align} 
If we condition the state on a successful run, the computational state will be: \[|\psi \rangle = \frac 12(\hat {I} + \exp [-\ii(\hat H - E_t)t_1])|\psi _i\rangle,\] where the probability of a successful outcome is encoded in the norm squared of the state $\langle \psi | \psi \rangle$.

Extending to a set of $N$ time samples $t_{j=1,\ldots, N},$ the post-RA state will be: \begin{equation}|\psi \rangle = \frac 1{2^N} \prod _{j=1}^N \left(\hat {I} + \exp[-\ii(\hat H - E_t) t_j]\right) |\psi _i\rangle.\end{equation}
\section{Cost Metric}
\label{appen:costmetric}
\subsection{Iterative Cost}
The advantages of the hybrid method are particularly apparent when iteratively constructing a many-body ground state. To simplify the setup, we consider a model defined on a 1D lattice of length $L = 2^\ell$. 
In the following discussion, $\ket {\psi _{\ell}}$ refers to the ground state of the model of size $L = 2^{\ell}$. Define $\kappa _\ell$ to be the total computational effort of constructing the many body state $\ket{\psi _\ell}$ out of $2^{\ell -1}$ copies of $\ket{\psi _1}$, which we assume are available at no cost.

\begin{lemma}
For $\ell \geq 2,$ the cost of preparing a system of size $2^\ell$ is related to the cost of preparing a system of size $2^{\ell -1}$ via:
\begin{equation}
    \kappa _{\ell} = \frac{T_{\ell } + B(\kappa _{\ell -1}, \tau _{\ell -1})}{p _{\ell}},
\end{equation}
where $\kappa_1\equiv 0$, $p_{\ell}$ is the probability of success of merging two systems of size $2^{\ell -1}$, $T_\ell $ is the total unitary evolution time for merging two systems of size $2^{\ell -1}$, $\tau _{\ell -1}$ is the unitary evolution time of a single preparation attempt for the subsystem, and $B(\kappa, \tau)$ is the ``both" function measuring the expected wait time for two independent machines to succeed.
\end{lemma}

\begin{proof}
    Consider having prepared two perfect copies of the state $\ket{\psi _{\ell -1}}$. By definition of the both function, this takes cost $B(\kappa _{\ell -1}, \tau _{\ell -1})$. To do the fusion step it will take time $T_{\ell}$, which includes the unitary evolution time for both adiabatic and Rodeo. Define $\tilde T_\ell \equiv T_{\ell} + B(\kappa _{\ell -1}, \tau _{\ell -1})$. Then, the expected cost of completing this step is the series:
    \begin{equation}
        \kappa _\ell = p_\ell \times \tilde T_{\ell} + p_\ell (1-p_\ell) \times 2\tilde T_{\ell} + p_\ell (1-p_\ell)^2 \times 3 \tilde T_\ell+\cdots = \sum _{j=1}^\infty p_{\ell} (1-p_{\ell})^{j-1} \times j\tilde T_{\ell} = \frac{\tilde T_\ell}{p_\ell},
    \end{equation}
    showing the Lemma. The special case of the first merging is included by the definition $\kappa _1 = 0$ where we assume the states $\ket{\psi _1}$ are free to prepare and take no computational cost to prepare. By the below definition, $B(0) = 0$.
\end{proof}

\begin{lemma}
    The ``both" function is:
    \begin{equation}
        B(\kappa, \tau) = \frac{\kappa (3\kappa - 2\tau)}{2\kappa - \tau},
    \end{equation}
    and defines the estimated continuous wait time for two machines to succeed, given they independently have an expected continuous cost $\kappa$ and a single-attempt duration $\tau$.
\end{lemma}

\begin{proof}
    We model the independent machines using the discrete-time Markov chain shown in Fig.~\ref{fig:both_function_markov_chain}, where the number in the circle indicates the number of succeeded machines. We define the expected number of discrete attempts required for one system to succeed as $N = \kappa/\tau$ (where $\kappa \geq \tau$). Assuming these attempts follow a geometric distribution, the success probability of a single attempt is $p = N^{-1} = \tau/\kappa$. 
    
    Let $E_0$ and $E_1$ be the expected number of additional attempts starting in state 0 and state 1, respectively. Using standard first-step analysis, we derive the following recursive state equations:
    \begin{equation}
        \begin{cases}
            E_1 = p^{-1}\\
            E_0 = 1 \times p^2 + 2 p(1-p) (E_1 + 1) + (1-p)^2 (E_0 +1)
        \end{cases}.
    \end{equation}
    By substituting $E_1$ into the second equation and solving for $E_0$, we isolate the expected number of discrete attempts for both to succeed:
    \begin{equation}
        E_0 = \frac{3 - 2p}{p(2 - p)}.
    \end{equation}
    Substituting $p = \tau/\kappa$ back into this result yields $E_0 = \frac{\kappa(3\kappa - 2\tau)}{\tau(2\kappa - \tau)}$. Finally, to convert this dimensionless attempt count into physical wait time, we multiply by the time per attempt $\tau$, yielding $B(\kappa, \tau) = E_0 \times \tau$, which proves the Lemma.
    \begin{figure}[h!]
        \centering
        \includegraphics[width=0.5\linewidth]{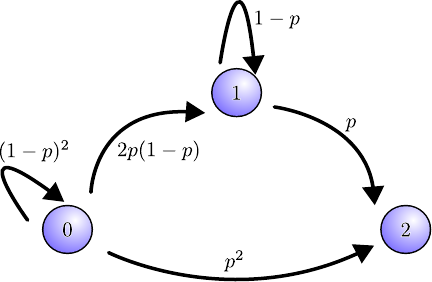}
        \caption{Both function Markov Chain}
        \label{fig:both_function_markov_chain}
    \end{figure}
\end{proof}

\subsection{Error Propagation}
The above cost formulas assume that the input states are perfect; however, this is not the case when building a many-body state from scratch. We hence need to consider how the errors add after sequential fusion steps. For simplicity, we assume that given a target infidelity of $\mathcal I$ of the final many-body state, we stop each fusion step at an intermediate infidelity of $\mathcal I_{\mathrm s}$. We show the following theorem:
\begin{theorem}
   Given a target infidelity of $\mathcal I$ to prepare the ground state $\ket{\psi _{\ell}}$ a system of size $L = 2^{\ell}$, each fusion step should be stopped at an intermediate infidelity of:
   \begin{equation}
       \mcalI = \frac{\mathcal I}{2^{\ell -1}-1}.
   \end{equation}
\end{theorem}

\begin{proof}
    Consider the intermediate step of merging two $\ket{\psi _{\ell-1}}$'s into a single $\ket{\psi _{\ell}}$. By the assumption that we stop fusion steps at $\mcalI$, the two halves of the system are not in perfect ground states but are in the state:
    \begin{equation}
        \tilde \rho _{\ell -1} = (1-\mcalI) \rho _{\ell -1}+ \mcalI \sigma _{\ell -1},
    \end{equation}
    where $\ket{\psi _{\ell -1}}\bra{\psi _{\ell -1}}$ is the ideal ground state while $\sigma _{\ell -1}$ is the error-space density matrix which we assume to be orthogonal to $\ket{\psi _{\ell -1}}$.

    The combined state on the two halves of the system reads:
    \begin{align}
        \tilde \rho _{\ell-1} \otimes \tilde \rho _{\ell -1} &= (1-\mathcal I_{\mathrm s}) ^2\rho _{\ell -1}\otimes \rho _{\ell -1}+ \mathcal I_{\mathrm s} (1-\mathcal I_{\mathrm s}) (\rho _{\ell -1} \otimes \sigma _{\ell -1} + \sigma _{\ell -1} \otimes \rho _{\ell -1}) + \mathcal I_{\mathrm s}^2 \sigma _{\ell -1} \otimes \sigma _{\ell -1} \\
        &= (1-2\mathcal I_{\mathrm s}) \rho _{\ell -1} \otimes \rho _{\ell -1} + \mathcal I_{\mathrm s} (\rho _{\ell -1} \otimes \sigma _{\ell -1} + \sigma _{\ell -1} \otimes \rho _{\ell -1}) + \mathcal O(\mathcal I_{\mathrm s}^2).
    \end{align}
    Consider now applying the fusion operation $\mathcal F(\cdot)$ to merge the two halves. 
    To evaluate the first term, note that the fusion operation is designed to map the perfect ground state of the unfused chain into the ground state of the fused chain plus some controlled error:
    \begin{equation}
        \mathcal F(\rho _{\ell -1}\otimes \rho _{\ell -1}) = (1-\mathcal I_{\mathrm s})\rho_{\ell}+ \mathcal I_{\mathrm s} \sigma _\ell .
    \end{equation}
    Next, consider the terms with one half in the ground state and the other in some error state (which is by definition an excited state of the chain). Both terms are some excited state of the unfused chain and are order $\mathcal O(\mathcal I_{\mathrm s})$. This means that the imperfect part of the fusion operation will produce a term $\mathcal O(\mathcal I_{\mathrm s}^2)$ and is hence ignored. We thus only consider the perfect part of the fusion operation. When operating perfectly, the first step of the fusion operation has no diabatic excitations, and hence results in an excited state of the length-$2^\ell$ chain, which is orthogonal to $\ket {\psi _\ell}$. But upon ideal operation, the subsequent Rodeo part of the fusion step will project out any orthogonal components to $\ket{\psi _\ell}$. Therefore the term only contributes $\mathcal O(\mathcal I_{\mathrm s}^2)$ terms and can be dropped. In sum:
    \begin{align}
    \mathcal F(\tilde \rho _ {\ell -1}\otimes \tilde \rho _{\ell -1}) &= (1-2\mathcal I_{\mathrm s}) [(1-\mathcal I_{\mathrm s})\rho _{\ell }+ \mathcal I_{\mathrm s} \sigma _\ell  ] + \mathcal O(\mathcal I_{\mathrm s}^2)\\
    &= (1-3\mathcal I_{\mathrm s}) \rho _{\ell} + \mathcal I_{\mathrm s} \sigma _\ell  + \mathcal O(\mathcal I_{\mathrm s}^2)
\end{align}
making the fidelity with the target state
\begin{align}
    \braket{\psi _\ell |\mathcal F(\tilde \rho _{\ell -1} \otimes \tilde \rho _{\ell -1})|\psi _\ell } &= 1-3\mathcal I_{\mathrm s}.
\end{align}

This can be extended inductively to show that infidelities accrue linearly. That is, upon $n_{\mathrm f}$ applications of fusion steps to prepare a many body state, the accumulated infidelity is $n_{\mathrm f} \times \mathcal I_{\mathrm s}$. To prepare state $\ket {\psi _{\ell}}$ from scratch, one must do $\sum _{\ell _i=1}^{\ell-1} 2^{\ell -\ell _i-1} = 2^{\ell -1}-1$ fusion steps.
\end{proof}

\section{Free Fermion Simulation}
\label{appen:freefermionsim}

Under the Jordan-Wigner transformation, the system maps into free fermions with the Hamiltonian:
\begin{equation}
    H = J \sum _{\braket{i,j}} (a_i^\dagger a_j + \mathrm{h.c.})
\end{equation}
allowing for efficient classical simulation.

The free-fermion simulation for the adiabatic fusion is straightforward, as the evolution operator is Gaussian and preserves the Slater determinant structure of the ground state. Across all system sizes, we observe a definitive late-time power-law scaling of the infidelity $\mathcal I\sim T^{-2}_{\mathrm r}$, while for short times the infidelity curve tapers out.
We hence fit the data to
\begin{equation}
    \mathcal I(T_\mathrm a) = \frac{\mathcal I(0)}{1 + (\tfrac {\mathcal I(0)}{A}) \times T_{\mathrm a}^2},
\end{equation}
where $A$ is the fitting parameter and the value tapers to $\mathcal I(0)$ as $T_{\mathrm a} \to 0$  and maintains the $\sim T_{\mathrm a}^{-2}$ scaling at late times. This fit slightly overestimates the performance in a small, early-time region by a negligible amount.

For the RA and hybrid methods, the RA evolution operator is a sum of two Gaussian operators, meaning that at each time step, we must branch into two separate evolutions since the state is not preserved as a Slater determinant. Given $r$ time steps/mid circuit measurements, the time complexity for a classical simulation thus becomes $\mathcal O(2^r)$. We hence must truncate all superiterations at some $r_{\mathrm{max}}$: We begin with a target Rodeo time $\tilde T_{\mathrm r}$ and $\alpha^{-1}$ common ratio of the series, constructing an infinite series with these parameters, and then truncating it to only include $r_{\mathrm {max}}$ terms.
The $\alpha$ is chosen by simulating small systems where we observe that a choice of  $ \alpha = 1 + 2.8 \times \frac{T_0}{T_{\mathrm r} L}$, where $T_0 = \pi/\Delta$ and $\Delta$ is the excitation gap. This choice of times is by no means optimal, but is sufficient to demonstrate the superiority of the hybrid method.
Both methods are fit with an exponential suppression function, where the rate of decay is fit to slightly underestimate the performance.
\bibliography{biblio}

\end{document}